\documentclass[12pt]{article}

\usepackage{scicite}
\usepackage{times}
\usepackage{graphicx}         %Include figure files
\usepackage{amsmath,amssymb}
\usepackage{notes2bib}

\newenvironment{sciabstract}{%
\begin{quote} \bf}
{\end{quote}}

\title{\textbf{The spectral dimension of human mobility}} 

\author
{Lei Dong$^{1,5 \dagger}$, Kevin O'Keeffe$^{1 \dagger}$, Paolo Santi$^{1,2\ast}$, Mohammad Vazifeh$^{1}$, \\
Samuel Anklesaria$^{1}$, Markus Schl\"apfer$^{3,4}$, Geoffrey West$^{4}$, Carlo Ratti$^{1}$ \\
\\
\normalsize{$^{1}$Senseable City Lab, Massachusetts Institute of Technology, Cambridge, MA 02139, USA}\\
\normalsize{$^{2}$Istituto di Informatica e Telematica del CNR, Pisa 56124, Italy}\\
\normalsize{$^{3}$Future Cities Lab, ETH Zurich, Zurich 8092, Switzerland}\\
\normalsize{$^{4}$Santa Fe Institute, Santa Fe, NM 87501, USA}\\
\normalsize{$^{5}$Institute of Remote Sensing and Geographical Information Systems,}\\
\normalsize{School of Earth and Space Sciences, Peking University, Beijing 100871, China}\\
\normalsize{$^\dagger$These authors contributed equally to this work.}\\
\normalsize{$^\ast$To whom correspondence should be addressed: psanti@mit.edu (P.S.).}\\
}

\begin{document} 
\baselineskip24pt
\maketitle

\begin{sciabstract}
\textbf{Human mobility patterns are surprisingly structured \cite{zipf1946p,erlander1990gravity,brockmann2006scaling,gonzalez2008understanding,song2010modelling,simini2012universal}. In spite of many hard to model factors, such as climate, culture, and socioeconomic opportunities, aggregate migration rates obey a universal, parameter-free, `radiation' model \cite{simini2012universal}. Recent work \cite{schlapfer2020hidden} has further shown that the detailed spectral decomposition of these flows -- defined as the number of individuals that visit a given location with frequency $f$ from a distance $r$ away -- also obeys simple rules, namely, scaling as a universal inverse square law in the combination, $rf$. However, this surprising regularity, derived on general grounds, has not been explained through microscopic mechanisms of individual behavior. Here we confirm this by analyzing large-scale cell -phone datasets from three distinct regions and show that a direct consequence of this scaling law is that the average `travel energy' spent by visitors to a given location is constant across space, a finding reminiscent of the well-known travel budget hypothesis of human movement \cite{mokhtarian2004ttb}. The attractivity of different locations, which we define by the total number of visits to that location, also admits non-trivial, spatially-clustered structure. The observed pattern is consistent with the well-known central place theory in urban geography \cite{christaller1933zentralen}, as well as with the notion of Weber optimality in spatial economy \cite{fujita2001spatial}, hinting to a collective human capacity of optimizing recurrent movements. We close by proposing a simple, microscopic human mobility model which simultaneously captures all our empirical findings. Our results have relevance for transportation, urban planning, geography, and other disciplines in which a deeper understanding of aggregate human mobility is key.}
\end{sciabstract}

\clearpage
%\section*{Introduction}
Individuals make regular visits to different places at a wide range of distance and visiting frequencies. This frequency depends on the type of activity performed at a destination locationn (eateries, shopping malls, work places etc) at a certain distance from an individual's origin place (often an individual's home location) \cite{christaller1933zentralen,zipf1949human}. In a recent study, we have shown that the number of visiting individuals follows an inverse square law of the production of frequency and distance \cite{schlapfer2020hidden}. More precisely, we can group the visitors to a given location $c$ by frequency of visitation $f$ during a reference period $T$, and consider the spectral flow rates $N_{c,f}(r)$: the total number of visitors who visit location $c$ from distance $r$ for $f$ times in $T$. The total number of individuals to $c$ is then $N_{total,c} = \sum_f \int N_{c,f}(r) r dr$. Here, we have computed $N_{c,f}(r)$ using datasets from three different regions: Greater Boston Area (the United States), Dakar region (Senegal), and Abidjan (Ivory Coast), see Supplementary Material (SM) and Table S1 for details. 

Following the approach in \cite{schlapfer2020hidden} and defining a high-resolution grid with cells of size $1km \times 1km$, we construct the user's movement in two main steps (see SM for details): $1)$ identify the {\em home cell} for each user, which we define as the grid cell where the user spent the most time at night (see Fig.1A-C); $2)$ for each (user, cell) pair, compute the number of monthly {\em visits} $f$, and {\em travel distance} $r$, from the home cell to the given cell by the given user, where a cell is considered visited if the user resides there for a minimum time of $\tau_{\text{min}} = 2~\text{hours}$. $r$ is defined as the geographical distance between the center of the user's home cell and the center of the visited cell. The desired $N_{c,f}(r)$ are then easily calculated from the data. 

\begin{figure*}[!htbp]
    \centering
    \includegraphics[width=1.\linewidth]{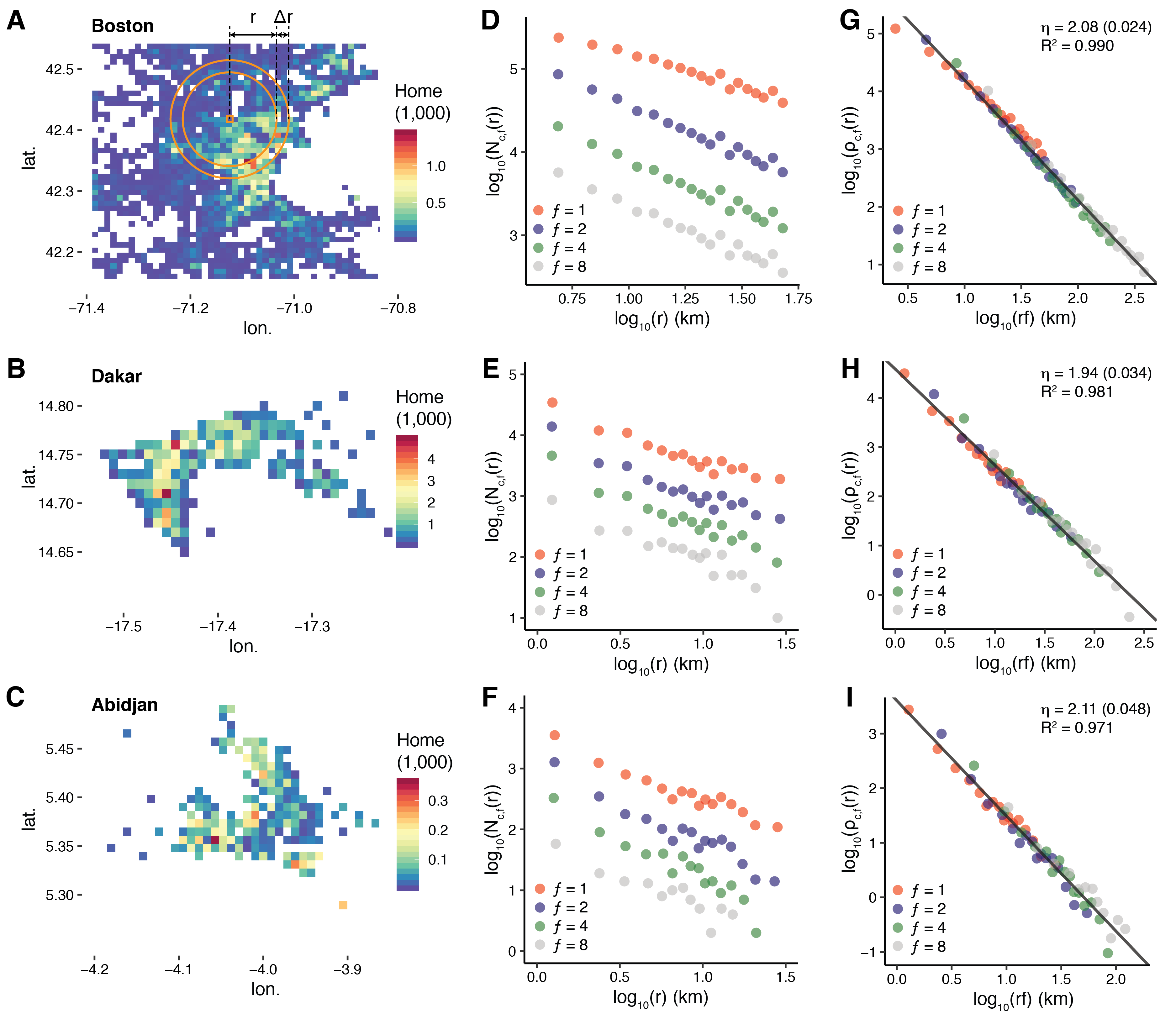}
    \caption{\textbf{Universality in the distance-frequency patterns of human movements.}  The home locations for Greater Boston Area (\textbf{A}), Dakar (\textbf{B}), and Abidjan (\textbf{C}). We show how $N_{c,f}(r)$ is calculated in (\textbf(A)). For a given cell, we count visitors from origin distance within $[r, r + \Delta r]$, see SM for details. (\textbf{D-F}) The number of visitors $N_{c,f}(r)$ making visits from distance $r$ averaged over a group of cells. Different values of visiting-frequency $f$ are shown in different colors. (\textbf{G-I}) Re-scaling of the same data with visiting-frequency, $f$. This confirms the prediction and analysis of ref.\cite{schlapfer2020hidden} which showed that the visit density for a center, $\rho_{c,f}(r)$, can be well-approximated by a single function $\rho_{c,f}(r) = \mu_{c}/(rf)^{-{\eta}}, \eta \simeq 2$, implying that the single parameter, $rf$, is sufficient to express the interplay between distance and the visiting-frequency, uncovered in ref.\cite{schlapfer2020hidden}. Here, data from Abidjan has been added to further confirm this result ($R^{2}s>0.97$ and standard errors of $\eta$s are shown in parentheses).}
    \label{fig:1}
\end{figure*}

Fig. 1D-F show different frequency groups -- hereafter called $f$-groups -- have different flow rates: for fixed travel distance $r$, $N_{c,f}(r)$ declines with $f$; the frequent visitors to a cell are outnumbered by the infrequent visitors. Strikingly, under the simplest transformation $r \rightarrow rf \, (n=1)$, the data collapse to a single, universal curve \cite{schlapfer2020hidden}, so that the visitation density from distance $r$ to a cell $c$, $\rho_{c,f}(r)$, can then be approximated as  $\rho_{c,f}(r) = N_{c,f}(r)/(2 \pi r) = {\mu_c / 2\pi (rf)^{-2}}$, where $\mu_c$ is a cell dependent `attractivity' measuring how popular a given cell is (Fig.~\ref{fig:1}G-I). This tells us that, in contrast to net migration rates \cite{zipf1946p,simini2012universal} -- which the gravity and radiation models endeavor to explain --, the main parameter governing {\em spectral} flow rates is not the distance $r$ but rather the product $rf$. Since it measures the total distance traveled by an individual during a given reference period, we interpret $E := r f$ as a {\em travel energy} (or alternatively, a {\em travel budget}). Our finding, then, is that the common structure between the spectral flow rates is the travel energy. Or put another way, though their radial distributions are different, the energy distributions of each frequency group ($f$-group) at a given cell are identical. Hence, $\rho_{c,f}(r) \propto \mu_c /(rf)^\eta = \mu_c / E^\eta$, where $\eta \approx 2$. 

A surprising consequence of this finding is that the average travel energy per visitor to a cell, $\langle E \rangle = E_{total} / N_{total}$, where $E_{total}$ is the total energy spent by visitors to a cell and $N_{total}$ is the total number of visitors, is spatially invariant – a kind of conservation law of human mobility:
\begin{equation}
    \langle E \rangle =  \frac{\sum_f \int_{r_{min}}^{r_{max}} r f N_{c,f}(r) 2\pi rdr}{\sum_f \int_{r_{min}}^{r_{max}} N_{c,f}(r) 2\pi rdr} =
     \frac{\sum_f \int_{r_{min}}^{r_{max}} (rf)^{-1} 2\pi rdr}{\sum_f \int_{r_{min}}^{r_{max}} (rf)^{-2} 2\pi rdr},
\end{equation}
\noindent

where $r_{min}, r_{max}$ are the minimum and maximum distances traveled by walkers in our datasets. We see the only cell dependent quantity, $\mu_c$, cancels out. Fig. 2 shows the conservation law is confirmed by our datasets. The spatial invariance of $\langle E \rangle$ is surprising because one might think that more attractive locations in a city would, on average, receive more travel energy from their visitors. In fact, more attractive places differ only in the number of visitors they receive, not the travel energy per visitor.

The spatial homogeneity of $\langle E \rangle$ led us to investigate the spatial distribution of the cell attraction parameters $\mu_c$. Recall these encode how popular, in terms of number of visitors, a given cell $c$ is. Fig. 3A shows $\mu_c$ for the Boston dataset have a clustered, spatial structure where the sizes of the clusters form a hierarchy. The emergence of clusters is expected: they form from the agglomeration effect of cities, -- that is, from the tendency of services and facilities to locate around city centers or sub-centers -- a finding consistent with the literature on urban structures \cite{christaller1933zentralen,anas1998urban,batty2008size,henderson2004handbook,bertaud2018order}, as well as previous empirical studies of urban mobility \cite{louail2014mobile,zhong2017revealing}. The emergence of the \textit{hierarchy} of cluster sizes is likely a result of another well known law of Zipf's \cite{zipf1949human}. To test this, we investigated if the cluster sizes are power law distributed. We used the City Clustering Algorithm (CCA) \cite{rozenfeld2008laws} to compute the clusters from data, which works as follows (see SM for details). First, the values of all cells with $\mu_c$ less than a threshold $\mu_c^*$ are set to zero. The values of all remaining cells are set to $1$. Second, the cells with value $1$ that are contiguous in space are merged recursively, until `islands' of $1$'s surrounded by $0$'s are formed, giving the desired set of clusters. Thus, given a threshold $\mu_c^*$, a set of clusters is generated. We chose the threshold $\mu_c^*$, by plotting the ratio of the area of the largest cluster to the sum of the areas of all the clusters formed in the Boston data for different $\mu_c^*$ (Fig. 3C). As seen, there is a critical value of $\mu_c^* \approx 10^2$ where the area ratio is minimized; this marks the onset of the emergence of a giant cluster and serves as a natural choice of $\mu_c^*$. Fig. 3D shows the distribution of cluster sizes at this $\mu_c^*$ do indeed follow Zipf's law \cite{zipf1949human}, a law fundamental in city science \cite{west2017scale}. We show a spatial plot of the clusters selected at $\mu_c^*$ in Fig. 3B. 

We now take stock of our findings: (i) the universal energy distribution and its associated conservation law, and (ii) the clustered spatial pattern of attractivity parameters $\mu_c$ whose size distribution match Zipf's law. Current models of human mobility cannot simultaneously account for both these observations. The popular exploration and preferential return model (EPR) \cite{song2010modelling}, which we will discuss shortly, accounts for (i) but not (ii) (Fig. S4). Here, building on the EPR model, we develop a model that can produce both (i) and (ii).

The EPR model is a random walk-like model. At each step with a certain probability, the walker chooses to explore a previously unvisited location via a L{\'e}vy jump \cite{zaburdaev2015levy}, namely, with a radial jump $\Delta r \sim (\Delta r)^{-1 - \alpha} $ and uniformly chosen angle $\theta \sim (2 \pi)^{-1}$. If the walker does not choose to explore she returns to a previously visited location with a certain probability (A detailed description of the EPR model is given in SM).

Notice the EPR model describes the motion of a single, independent walker: in a population of walkers following the EPR model, the individuals do not interact. In reality, however, individuals' motions \textit{do} interact \cite{strandburg2015shared}: the motions are correlated through common attraction points and activity hubs. That is, people do not choose destinations that are entirely independent of other peoples' destinations; they tend to visit `popular' places -- places visited frequently by \textit{other} people. Thus, in ignoring this coupling between walkers' motion, the EPR model is unable to reproduce observation (ii): the clustered distribution of attractivity parameters $\mu_c$. As shown in Fig. S4, the EPR model's $\mu_c$ are uniform across space, in stark contrast to real data (Fig. 3A).

To account for clustered $\mu_c$, we introduce the notion of \textit{preferential exploration}, resulting in a modification of the EPR model that we call preferential exploration and preferential return (PEPR). Preferential exploration is achieved by coupling the walkers' motion. When exploring a new location, a walker is preferentially attracted to popular places, i.e., places visitors have spent large amounts of energy getting to. The radial jump distances $\Delta r$ are still sampled from $P(\Delta r) \sim (\Delta r)^{-1 - \alpha}$ but the angle $\theta$ the walker chooses to jump in is no longer drawn uniformly at random. Instead, angles which correspond to regions of high visitation are selected preferentially. Let, as before, $E_{total}$ be the aggregate energy spent getting to the cell by all visitors to that cell. Further, let the \textit{diffused aggregate energy} $E_c(\theta; R)$ of cell $c$ be the sum of the aggregate energy of all cells within distance $R$ of $c$ between angles $\theta$ and $\theta + d \theta$. Then walkers following the PEPR model sample $\theta$ from $ P(\theta; R,\nu) \sim E_c(\theta;R)^{\nu}$. We show a schematic of the PEPR model in Fig. 4A. 

Figs. 4BC show the PEPR model reproduces finding (i), the spectral flow rates and their scaling collapse, and more importantly finding (ii), realistic hierarchical visitation patterns: a qualitatively similar spatial pattern of clusters (Fig. 4D) and a quantitatively accurate cluster size distribution (Fig. 4F). Regarding the spatial patterns, we say ``qualitatively similar'' since the exact layout of the model clusters is different to that of real data. For example, in the real data there is a large cluster located on the coast (corresponding to Boston city) surrounded by multiple smaller clusters, which is different to the simulation data (Fig. 4D). Reproducing the clustered spatial patterns at this level of accuracy is however beyond the scope of the PEPR model since it ignores many complexities which likely influence the development of human towns/cities such as natural resources, rivers, topography, etc (see SM). Furthermore, the PEPR model was run on a square lattice, whereas Boston has an irregular geometry.

Our results support the well-known Central Place Theory \cite{christaller1933zentralen} of urban science which to date is (at large-scale) empirically unsupported. The theory asserts that `urban centers' form an orderly hierarchy arranged in space, where larger centers, which provide more `high-level' services (e.g., shopping centers, museums, theaters), are surrounded by smaller centers, which provide `local-level' services (e.g., groceries, primary schools, clinics). The rationale behind the theory is that such an arrangement minimizes the total distance traveled by the population, and is in that sense optimal. Our work corroborates both aspects of Central Place Theory: the clustered spatial pattern of $\mu_c$ we observed (Fig. 3) is consistent with the hierarchical structure, and in SM we show the conservation law $\langle E \rangle = const$ across space accords with the minimum-distance optimality. In addition, we show that the average distance traveled by individuals for a given visiting frequency $\langle r \rangle_f$ obeys the relation $\langle r \rangle_f = K / f$, which also serves as a validation of the Central Place Theory (Fig. S5).

Central Place Theory is rooted on an individual-level least-effort principle \cite{zipf1949human}, and an emerging \textit{self-organized optimality} \cite{batty2013new}. To strengthen the evidence for this intriguing possibility, we computed the Fermat-Toricelli Weber \cite{weber1929alfred} metric of our dataset. This is a metric used in spatial economy to quantify optimality from the perspective of the activity centers in a city (buildings, shops etc). Each cell $c$ is assigned an index $ \Delta{\mathcal{D}}_{\text{total}} / \mathcal{D}_{total} \in [0,1]$, where $\mathcal{D}_{total}$ is the total distance traveled by the reference population that visits $c$, and $\Delta{\mathcal{D}_{total}}$ is the improvement in overall distance traveled by the reference population gained by relocating the destination cell to another position on the grid. If the location of a cell is already optimal for the reference population, $\mathcal{D}_{total}$ cannot be reduced by relocating that cell and therefore the index is $0$. If the location of the cell is suboptimal, the index is close to 1. Remarkably, Fig. S6 shows most cells in our Boston dataset are close to their ``Weber optimal'' locations, having $\Delta{\mathcal{D}}_{\text{total}} / \mathcal{D}_{total} \approx 0 $. We give a full account of FTW theory and our computations in SM.  

This study provides evidences of self-organized optimality of a human collective behavior, namely, day-to-day mobility. In contrast, many results in game theory show that collective behavior is non-rational and far from the socially desired outcome \cite{kreps1990game,myerson2013game}. This non-rationality is thought to be due to cognitive limitations, that is, from the inability of the human mind to completely understand the complex system in which the human operates \cite{brubaker2013limits}. The results of this study stand as a clear \textit{counter} example to this. They demonstrate that collectively, humans are able to overcome their cognitive bounds and achieve optimal group-level behavior --  an important and hopeful finding for the human mind.

%\bibliography{ref.bib}
%\bibliographystyle{Science}

\subsection*{Acknowledgments} We thank W.P. Cao for assistance to perform the CCA analysis, and all the members of the MIT Senseable City Lab Consortium for supporting this research. L.D. acknowledges funding from the National Natural Science Foundation of China (No. 41801299).

\subsection*{Author contributions} C.R., P.S., G.W., L.D., and K.O. designed the research. L.D., K.O., and P.S. performed the research. L.D., K.O., M.V., S.A., and M.S. analyzed data. L.D., K.O., P.S., and M.V. constructed the model and wrote the paper. All authors reviewed the paper.

\subsection*{Competing interests} The authors declare no competing interest.

\subsection*{Data and code availability}
The data and code to replicate this research can be requested from the authors.

\clearpage

\begin{figure*}[!htbp]
    \centering
    , whi\includegraphics[width=1.\linewidth]{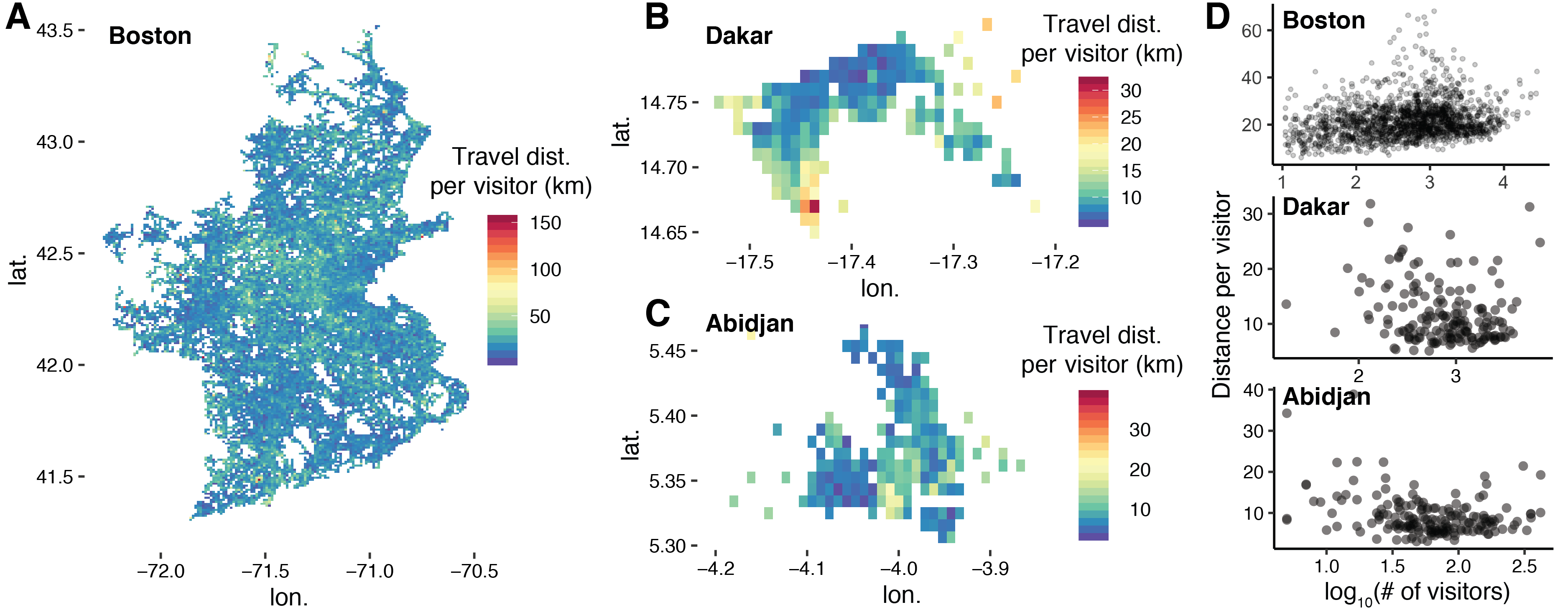}
    \caption{\textbf{Constant travel energy per visitor.} (\textbf{A-C}) The average energy $\langle E \rangle$ spent by an individual to visit a cell manifests uniformity across space consistent with the notion of travel budget per visitor as discussed in the paper. Note that the southern part of Dakar is an important port for Senegal, thus a lot of non-local visitors travel to this place, making the travel distance higher than the remaining places (but still within the same order of magnitudes). (\textbf{D}) The scatter plots of number of visitors and travel distance per visitor. The $R^2$s of linear regression between number of visitors and distance per visitor are very small (Greater Boston Area, $R^2 = 0.0167$, n = 14,273, $p$-value $<$ 0.005; Dakar, $R^2 < 0.001$, n = 173, $p$-value = 0.996; Abidjan, $R^2 = 0.005$, n = 183, $p$-value = 0.355).}
    \label{fig:2}
\end{figure*}

\begin{figure}[!htbp]
    \centering
    \includegraphics[width=.8\linewidth]{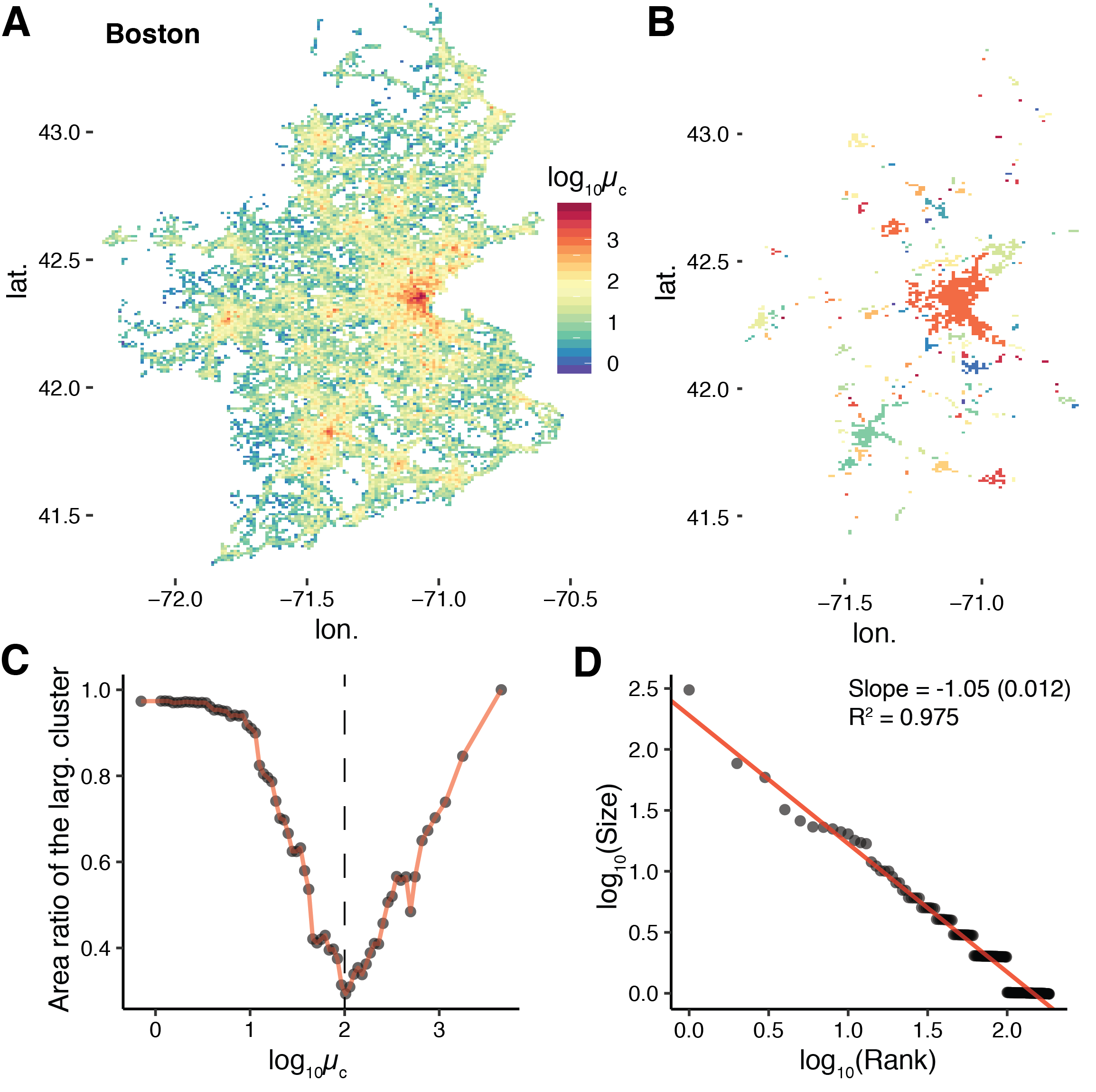}
    \caption{\textbf{Hierarchical structure of attractiveness, $\mu_c$.} (\textbf{A}) Geographical pattern of $\mu_c$ in Greater Boston Area. We derive cell specific $\mu_c$ by fitting Eq. (1) with the ordinary least squares regression. We set different thresholds $\mu_c^*$ for $\mu_c$ and then use the City Clustering Algorithm proposed in \cite{rozenfeld2008laws} to derive the continuous clusters with $\mu_c$ over the threshold (\textbf{B}). We calculate the area ratio of the area of the largest cluster to the sum of the areas of all clusters (\textbf{C}), derive the coefficient of the rank-size distributions at the critical value of $\mu_c^{*} \approx 10^2$ (vertical dashed line in (\textbf{C})), and present the detected clusters in (\textbf{B}) with different colors. When $\mu_c$ is very small, the whole Greater Boston Area would be connected to a single cluster, resulting in the area ratio $\approx 1$. When $\mu_c$ is very large, only one cluster (Boston downtown) would exist, also resulting in the area ratio $\approx 1$. (\textbf{D}) Statistical summary of the rank-size regression at the critical value of $\mu_c^{*}$: slope = -1.05 (0.012), $R^2 = 0.975$, indicating a well-fitted Zipf's law.}
    \label{fig:3}
\end{figure}

\begin{figure*}[!htbp]
    \centering
    \includegraphics[width=1.\linewidth]{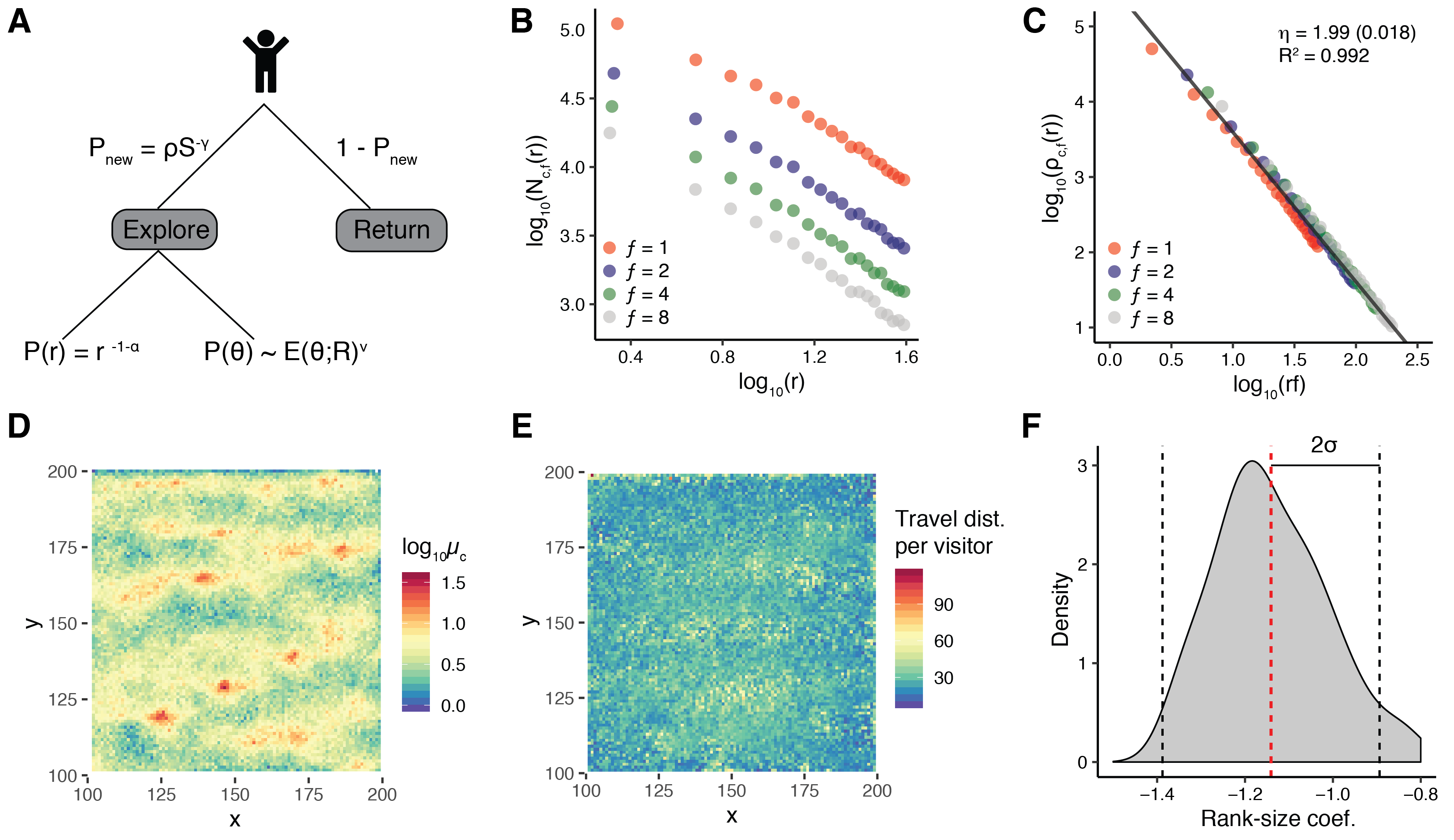}
    \caption{\textbf{PEPR Model and simulation results.} (\textbf{A}) Schematic of the PEPR model. (\textbf{B-F}) Simulation results on a lattice (with parameters $\alpha=0.55$, $\rho=0.6$, $\gamma=0.21$, $R=10$, $\nu=4$, and the number of agents = $1 \times 10^5$). (\textbf{B}) Relations of the number of visitors $N_{c,f}$ and $r$ with different $f$. (\textbf{C}) Similar to Fig. 1G-I we rescale (\textbf{B}) with visiting-frequency $f$, and all data points collapse onto a straight line ($\eta \simeq 2$, $R^2 = 0.992$). (\textbf{D}) Attractiveness $\mu_c$ generated by our model shows some significant spatial clusters. (\textbf{E}) The energy landscape based on the simulation results, which support the constant energy hypothesis in Eq. (2). (\textbf{F}) We repeat 50 simulations with the same parameters, and calculate the coefficient of the rank-size distribution at the critical value of $\mu_c$ ($\mu_c^* = 10$). The mean value of the coefficient is -1.14 (the red dashed line) and the 95\% confidential interval is [-1.39, -0.894] (the black dashed lines), showing a well-fitted Zipf's law, which is also similar to the empirical finding (Fig. 3D).}
    \label{fig:4}
\end{figure*}

\clearpage

\section*{Supplementary Materials}

\begin{itemize}
    \item Materials and Methods
    \item Tables S1
    \item Figures S1-S9
\end{itemize}

\section*{Materials and Methods}

\subsection*{Boston data}
Individuals' movements in Greater Boston Area are inferred from mobile phone Call Detailed Records (CDR) data collected over a span of 4 months. The dataset is provided by a company, and has been used in our previous studies \cite{ schlapfer2020hidden}. The raw data contains about 2 million anonymized users.

\subsection*{Dakar data}
The Dakar dataset is based on anonymized Call Detailed Records (CDR) provided by the Data for Development (D4D) Challenge. The detailed information of this dataset is provided in \cite{de2014d4d}. Here, we use the SET2, which includes individual trajectories for 300,000 sampled users in Senegal, and after the preprocess, we have 173,000 users and 173 cells in Dakar region during two weeks of January, 2013. We also use the datasets of March, June, and August of 2013 to verify the robustness of the observed universality of distance-frequency, see Fig. S3.

\subsection*{Abidjan data}
The Abidjan dataset is also based on anonymized CDR provided by the D4D Challenge. The structure of the data and the data preprocessing method are detailed in \cite{blondel2012data}. Here, we use the SET2 of the original dataset. It contains individual trajectories for 50,000 random sampled users in the Ivory Coast, and after the preprocess, we have 18,000 users and 183 $1km \times 1km$ cells in Abidjan during two weeks of December, 2011.

\subsection*{Data preprocessing}
CDR are generated only for voice calls, text messages or data exchanges and therefore have limited resolution in time. The geographic location of the cell towers and their density determines the accuracy of location measurements through triangularization techniques. Therefore, the trajectories extracted from CDRs constitute a discrete approximation of the moving population $M(x; y; t)$. There are several steps in preprocessing of the data before it can be suitable for use in our analysis. 

The main steps are: i) Partitioning of the study area. The area under study is partitioned into a rectangular grid. ii) For each grid cell of size $1km \times 1km$, we identify the individuals that have visited the location with a given frequency $f$, for instance $f = 5$ distinct days in a month for Boston (or bi-weekly for Dakar and Abidjan), while staying there for a minimum time $\tau_{min} = 2{\text h}$. Performing a robustness analysis shows that the result of our study is not sensitive to small changes in $\tau_{min}$. iii) For each person, we determine the home location as the grid cell which has been visited during most nights, i.e. between $7$pm and $7$am of local time. By summing over all days in a given time window (one month for Boston, and two weeks for Dakar and Abidjan), one can find the home cell with high level of confidence for the majority of subjects. The resident population $P_i$ of a given cell $i$ is then defined as the the total number of assigned persons to that cell. The number of visitors for each cell is defined as the total number of distinct, non-resident individuals visiting that cell. The number of visits for each cell is the total number of times that cell has been visited during the time window of interest. In Fig. S1, we present the visitation distributions for Boston, Dakar and Abidjan, respectively.
 
The duration of stay criterium on defining cell visits yields a list of cells visited by that subject during a day. By aggregating those visits over the course of a month (or two weeks) for each subject, we obtain a visiting-frequency vector of dimension $N_{\text cells}$ which is equal to the number of cells on the geographical grid. The $i$-th component of this vector represents the number of times the $i$-th cell has been visited by that subject. We then construct the overall visit matrix for each month $\mathcal{M}$. The $ij-$component of this matrix is the number of times $j$-th cell have been visited by the $i$-th subject. Although this matrix is huge in dimensions, its sparseness allows fast computation to derive various aggregate mobility related measures.
 
Here, the distance between cells is calculated by the haversine formula, which derives the great-circle distance between two points on a sphere. To count the number of visitors that cell $c$ received from origin distances $[r, r + \Delta r]$, we take $\Delta r = 2km$ for Boston, and $\Delta r = 1km$ for Dakar and Abidjan as the latter two regions are much smaller compared with Boston area. Meanwhile, to reduce the noise of the `tail' part of the aggregated visit, we take log-bins for distances over $20km$ in Boston dataset and over $10km$ in Dakar and Abidjan datasets (Fig. 1 D-F).

\subsection*{Quantifying spatial structure} 
We use City Clustering Algorithm (CCA) to derive spatial clusters of cell attractiveness. CCA, proposed in \cite{rozenfeld2008laws,rozenfeld2011area}, defines a `city' as a maximal, spatial continuous area with granular population data. The algorithm takes three steps: First, set a population threshold $P_{*}$ and binarize the study area into 0, 1 values -- cells with population over $P_{*}$ are set to be 1, otherwise to be 0. Second, the algorithm picks a populated cell (value = 1) randomly and adds the nearest populated cells recursively until all the nearest neighbors are unpopulated cells (value = 0). Third, repeat the picking and merging process until all populated cells belong to one specific cluster. This method is intuitive and can divide the US metro area into different clusters as shown in \cite{rozenfeld2011area}. 

In fact, CCA is not limited to use population as the input layer. However, no matter what kind of input layers used to perform CCA, the common problem is finding the proper threshold $P_{*}$ to binarize the urban area. A recent study proposes to employ percolation theory to solve the parameter selecting problem of CCA \cite{cao2019quantifying}. The paper has demonstrated that tuning the threshold $P_{*}$, a giant cluster would emerge as $P_{*}$ reaches a certain point in datasets of population, nighttime light, and road networks, which is in line with the two-dimensional percolation process \cite{cao2019quantifying}. We also find similar behavior when tuning the threshold of attractiveness $\mu_c$ in our case, which is likely to reflect the self-organization nature of urban systems. By setting $\mu_c$ at the critical value and performing the CCA, we have a giant cluster and a large number of smaller ones (Fig. 3B). To test the Zipf's law, we then run the ordinary least squares (OLS) regression between the cluster size and its corresponding rank among all detected clusters:

\begin{equation}
    \log_{10} Size_{i} = \beta_0 + \beta_1 \log_{10} Rank_{i} + \epsilon_{i}~,
\end{equation}
\noindent
where $Rank_{i}$ is the size rank of cluster $i$. We derive the parameter of interest $\beta_1$, and report the regression results in the main text (Fig. 3D). Zipf's law is considered to be a rank-size distribution of $\beta_1 = -1$.

\subsection*{Model and simulation}

\subsubsection*{EPR model}

The EPR model is a random walk-like model. At each step, the walker decides whether to explore a new, previously unvisited location with probability $P_{\text{new}}=\rho S^{-\gamma}$ where $S$ is the number of locations she has visited so far and $\rho$, $\gamma$ are model parameters. If the walker decides to explore, she jumps a distance $\Delta r$ sampled from $P(\Delta r) \propto |\Delta r|^{-1-\alpha} $ -- $\alpha$ is another model parameter -- at an angle $\theta$ chosen uniformly at random $P(\theta) = (2 \pi)^{-1}$ (i.e., does a L{\'e}vy flight \cite{zaburdaev2015levy}, to make the jump sizes consistent with empirical data \cite{barbosa2018human}). But if the walker does not choose to explore (which occurs with probability probability $1-P_{\text{new}}$) she returns to a previously visited location with probability proportional to the number of previous visits in each location. 

\subsubsection*{PEPR model and simulation}
We simulated $1 \times 10^5$ agents moving according to the model's rules on a $300 \times 300$ grid of cells. Home locations for these agents were assigned uniformly at random across the grid. Analysis was performed only on the $100 \times 100$ center region to eliminate boundary effects. The model parameters for data shown in Fig. 4BC were found by strobing over a grid in parameter space and selecting the parameters which led to the desired scaling collapse (exponent 2). $\alpha$, $\rho$, and $\gamma$ used here are also consistent with empirical findings~\cite{song2010modelling}. Similarly, those from Fig. 4D-F were those which led to a cluster size distribution which followed Zipf's law. 

\subsection*{Weber equilibrium}
We here analyze the role places of particular importance play in the formation of self-organized patterns of urban settlements by analyzing the efficiency of these cells from the point of view of spatial-economic theory. According to this theory, the urban population distribution is driven by centrifugal and centripetal forces which exist due to the economic competition in minimizing the transportation costs to important resources. We can investigate the efficiency of attractor cells quantitatively, by studying how close the total transportation cost of incoming visits to each cell is to the optimum transportation cost defined as the minimum possible value for the total transportation distance from visitor’s home-location. This problem can be formulated as the Fermat-Torricelli-Weber (FTW) problem on a square grid.

We define a bi-directed visit OD-flow spatial network in which the nodes correspond to geographical cells and the directed edges are weighted according to the number of visits exchanged between pairs of nodes. The visits flow matrix is an asymmetric square matrix which contains all the information about the visiting patterns and is defined as

\begin{equation}
\mathcal{V}_{ij} = \text{ total visits from} \; \mathcal{C}_i \;\text{to}\; \mathcal{C}_j
\end{equation}

Note that in general $\mathcal{V}_{ij} \neq \mathcal{V}_{ji}$. 

In the Weber problem in location theory, the optimal point is a point which minimizes the total distance from $n$ points on a plane -- Fig. S6. One can consider this problem on a grid where the optimum location can be chosen from a finite number of points corresponding to the centre of cells on a geographical grid, and the optimum is a cell which minimizes the overall transportation distances from where the visits originate from.

We define the Weber matrix as follows: 

\begin{equation}
\mathcal{W}_{ij} = T[\mathcal{C}_i\rightarrow \mathcal{C}_j]
\end{equation}

where $T[\mathcal{C}_i\rightarrow \mathcal{C}_j]$ is the total distance travelled by visitors of $i$-th cell if this cell where placed in the location of $j$-th cell. Using the distance matrix $\mathcal{D}$ and the visits flow matrix $\mathcal{V}$ we can compute the Weber matrix

\begin{equation}
\mathcal{W}_{ij} = \sum_{k} \mathcal{D}_{j k} \mathcal{V}_{k i} = [{\mathcal{D} \cdot \mathcal{V}}]_{ji}
\end{equation}

Each row of the Weber matrix contains all the possible values of the objective function defined according to the Weber problem on a grid for the corresponding cell. The question is how close the value of the actual total transportation distance for each cell, which correspond to the value on the diagonal axis of the matrix, is to the minimum possible value for each row, corresponding to the FTW cell. One way to measure this closeness is to see how much improvement can be gained for each cell if we move each cell to its FTW location. We define the fractional improvement as the ratio of the total energy improvement gained for each cell to the actual energy which is given as

\begin{equation}
\frac{\Delta \mathcal{D}_i}{\mathcal{D}_i} = \frac{ \mathcal{W}_{ii} - \text{min}({\mathcal{W}_{i*}})}{\mathcal{W}_{ii}}
\end{equation}

The above quantity is always between zero and one. The value zero corresponds to the extremum case where no improvement can be gained, meaning that the cell's location coincides with the optimal transportation location. The value would be equal to one when the transportation distance can be reduced to zero by moving the cell. Since the number of visits a cell gets does not change as we relocate the cell, the fractional distance per visit improvement, i.e., $\frac{\Delta \mathcal{D}_{iv}}{\mathcal{D}_{iv}}$ is equal to the fractional total distance improvement,  

\begin{equation}
\frac{\Delta \mathcal{D}^{\text{per visit}}_{i}}{\mathcal{D}^{\text{per visit}}_{i}} = \frac{\Delta \mathcal{D}_{i}}{\mathcal{D}_{i}}
\end{equation}

The average distance per visit can be quite large yet the highly important cells are very close to their FTW cell in the majority of the cases. In Fig. S6 we plot the total received visits by each cell versus the fractional improvement which can be gained by relocating them to their FWT point. As seen from the figure, the majority of the highly visited cells have low fractional improvement. The exceptions to this pattern are the few cells in the yellow box in Fig. S7. To see why these cells were exceptional, we checked the location of the cells on the map and found that in the majority of cases, they correspond to tourism attraction points near beaches, lakes, etc, which explains why they are anomalous -- these locations having an intrinsic reason to be located where there are, as opposed to being there so as to optimize their FTW score.

%%%%%%%%%%%%%%%%%%%%%%%%%%%%%%%%%%%%%%%%%%%%%%%%%%%%%%%%%%%%%%%%%%%%%%%%%%

\subsection*{Topography}
In the main text we showed that the spatial pattern of the visitation rates of the PEPR model (Fig. 4D) were different to those of the real Boston data (Fig. 3A). This mismatch was not surprising since the PEPR model only models how the interactions between individuals influences their movements and is blind to the terrain on which the individuals move. Presumably, different types of terrain would attract / repel people with different strengths. For example, areas with lots of natural resources would naturally attract settlement, as would rivers and coasts be attractive since they influence trade. Put generally, human movement would be influenced by topography, an effect which the PEPR model does not strive to capture.

A thorough study of the role of topography in human movement is beyond the scope of the present work. We here however take a first step in this direction by running the PEPR model on a non-trivial geometry to see if it leads to more realistic spatial visitation patterns. In the main text, the PEPR model was run on a square lattice, a crude approximation of the irregular geometry of Boston (Fig. 3A). In Fig. S8 we show the visitation pattern of the PEPR model when run on a lattice with a simple perturbation: a rectangular chunk removed from one side. As seen, the spatial visitation pattern is not qualitatively altered, demonstrating that other topographical features are needed to recover the hierarchical pattern observed in real data (Fig. 4D main text). 

%%%%%%%%%%%%%%%%%%%%%%%%%%%%%%%%%%%%%%%%%%%%%%%%%%%%%%%%%%%
\clearpage
\begin{table*}[!htbp]
    \centering
    \begin{tabular}{c|c|c|c|c|c}
    \hline
    \hline
        Region & Country & \# of CDR users & CDR date & Population & Area\\
        & & after preprocess & & $million$ & $km^2$\\
    \hline
        Greater Boston Area & US & 340,000 & 2009 & 4.73 & 11,700 \\
        Dakar Region & Senegal & 173,000 & 2013 &  2.96 & 547\\
        Abidjan & Ivory Coast & 18,000 & 2011-12 & 4.70 & 422\\
    \hline
    \hline
    \end{tabular}
    %\caption{}
    \begin{flushleft}
    \textbf{Table S1 $|$ Statistical summary of three regions.} Population of Abidjan was derived from 2014 census of Ivory Coast. Population and area data of Dakar region were derived from 2013 census of Senegal. Population and area data of Greater Boston Area were derived from Wikipedia.
    \end{flushleft}
    \label{tab:s1}
\end{table*}

\clearpage
%%%%%%%%%%%%%%%%%%%%%%%%%%%%%%

\begin{figure*}[!ht]
    \centering
    \includegraphics[width=1.\linewidth]{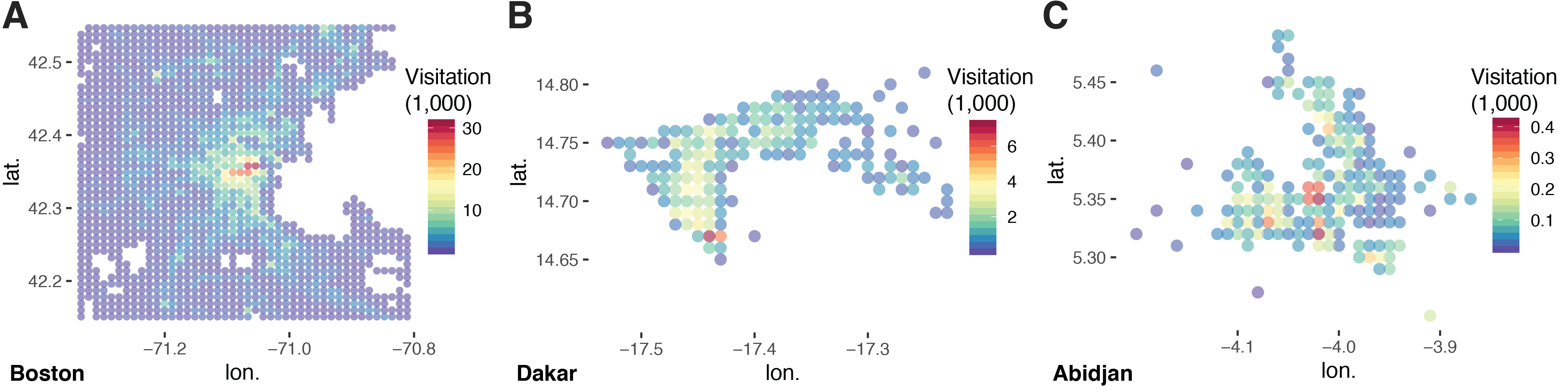}
    %\caption{}
    \begin{flushleft}
    \textbf{Fig. S1 $|$ Geographical distributions of the total visitation.} \textbf{A}, Boston; \textbf{B}, Dakar; and \textbf{C}, Abidjan.
    \end{flushleft}
    \label{fig:sfig1}
\end{figure*}

\clearpage
\begin{figure*}[!ht]
    \centering
    \includegraphics[width=1.\linewidth]{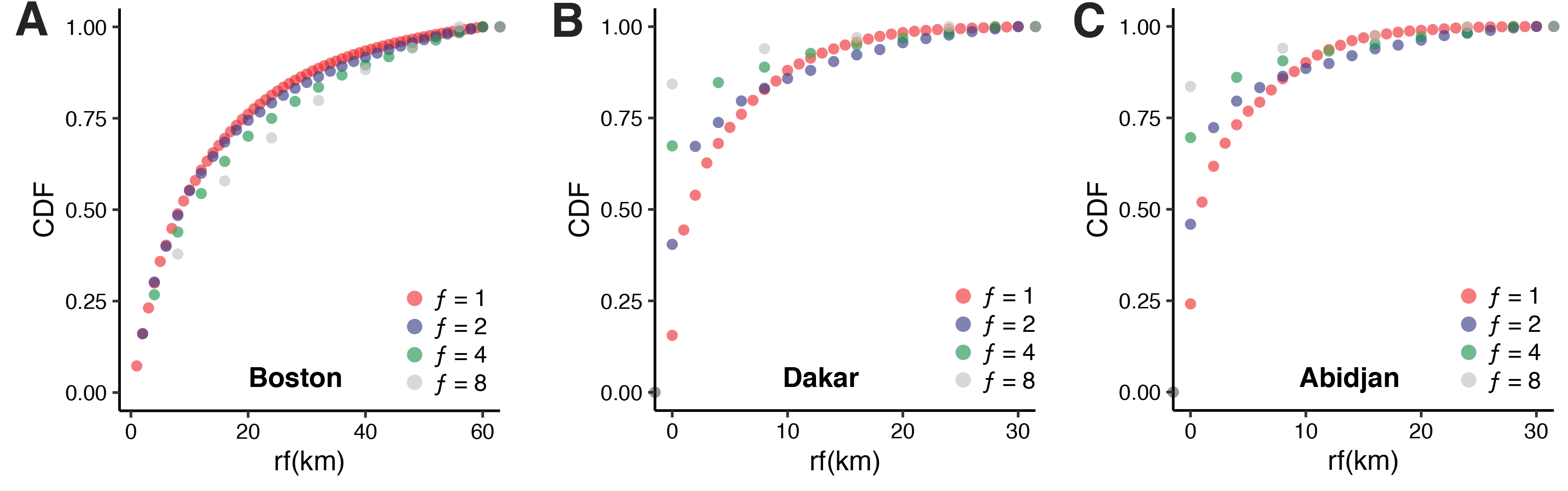}
    %\caption{}
    \begin{flushleft}
    \textbf{Fig. S2 $|$ The cumulative distribution of the number of visitors visiting from distance within $r$ radius for various visiting frequencies $f$.} \textbf{A}, Boston; \textbf{B}, Dakar; and \textbf{C}, Abidjan. The curves for different frequencies collapse approximately into a single curve after rescaling the distance with frequency, $r \rightarrow rf$.
    \end{flushleft}
    \label{fig:sfig2}
\end{figure*}

\clearpage
\begin{figure*}[!ht]
    \centering
    \includegraphics[width=1.\linewidth]{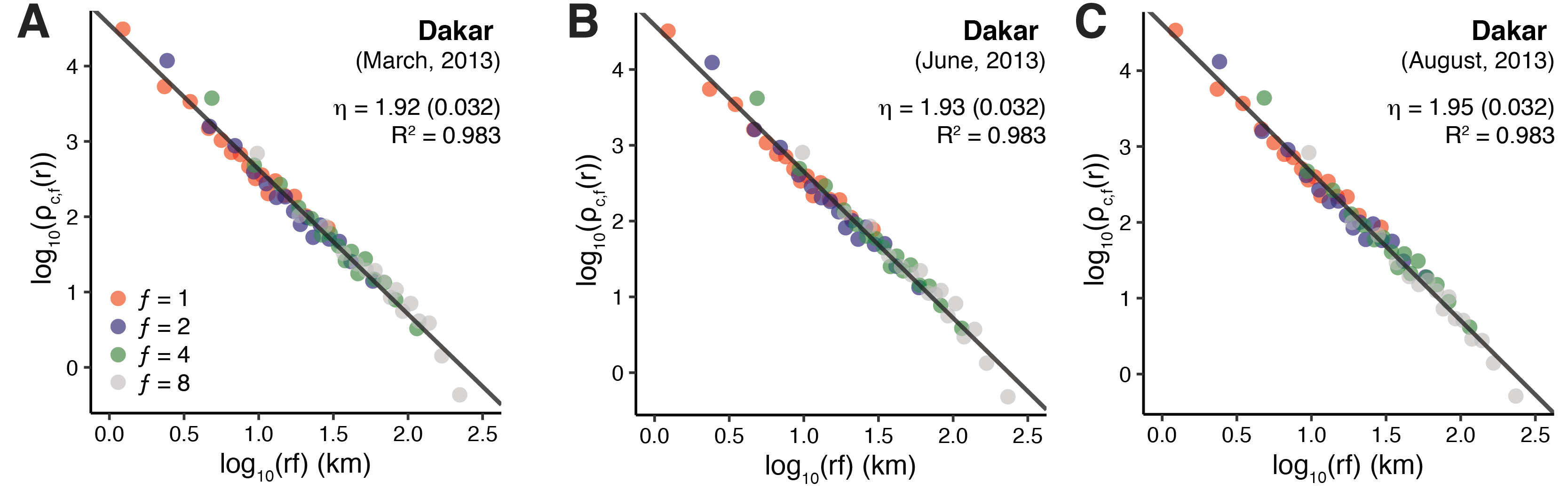}
    %\caption{}
    \begin{flushleft}
    \textbf{Fig. S3 $|$ Robustness check of the universal $rf$ in different time period.} Dakar datasets of \textbf{A}, March, 2013; \textbf{B}, June, 2013; and \textbf{C}, August, 2013. The variation density for a wide range of $r f$ can be well-approximated by a single function $\rho_{c,f}(r) = \mu_c \cdot (rf)^{-{\eta}}, \eta \simeq 2$ ($R^2 > 0.98$).
    \end{flushleft}
    \label{fig:sfig3}
\end{figure*}

\clearpage
\begin{figure*}[!ht]
    \centering
    \includegraphics[width=.85\linewidth]{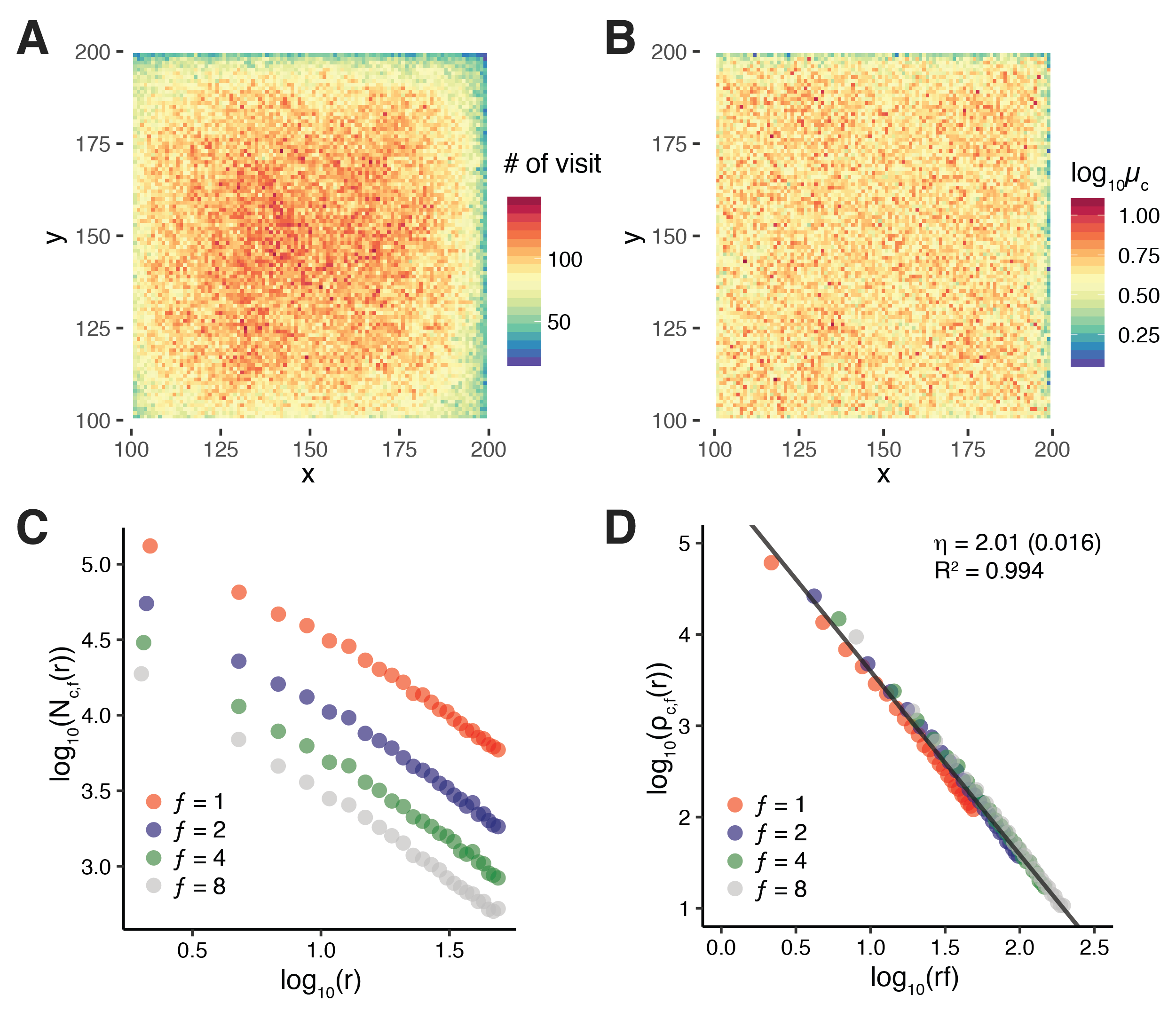}
    %\caption{}
    \begin{flushleft}
    \textbf{Fig. S4 $|$ Simulation results of the EPR model}. Visitations (\textbf{A}) and attractivity parameters $\mu_c$ (\textbf{B}) generated by the EPR model are uniform across space, which is in contrast to real data (Fig. 3 and Fig. S1). \textbf{C, Dd}, Similar to Fig. 4B, C, EPR model can reproduce Eq.(1).
    \end{flushleft}
    \label{fig:sfig4}
\end{figure*}

\clearpage
\begin{figure*}[!ht]
    \centering
    \includegraphics[width=.75\linewidth]{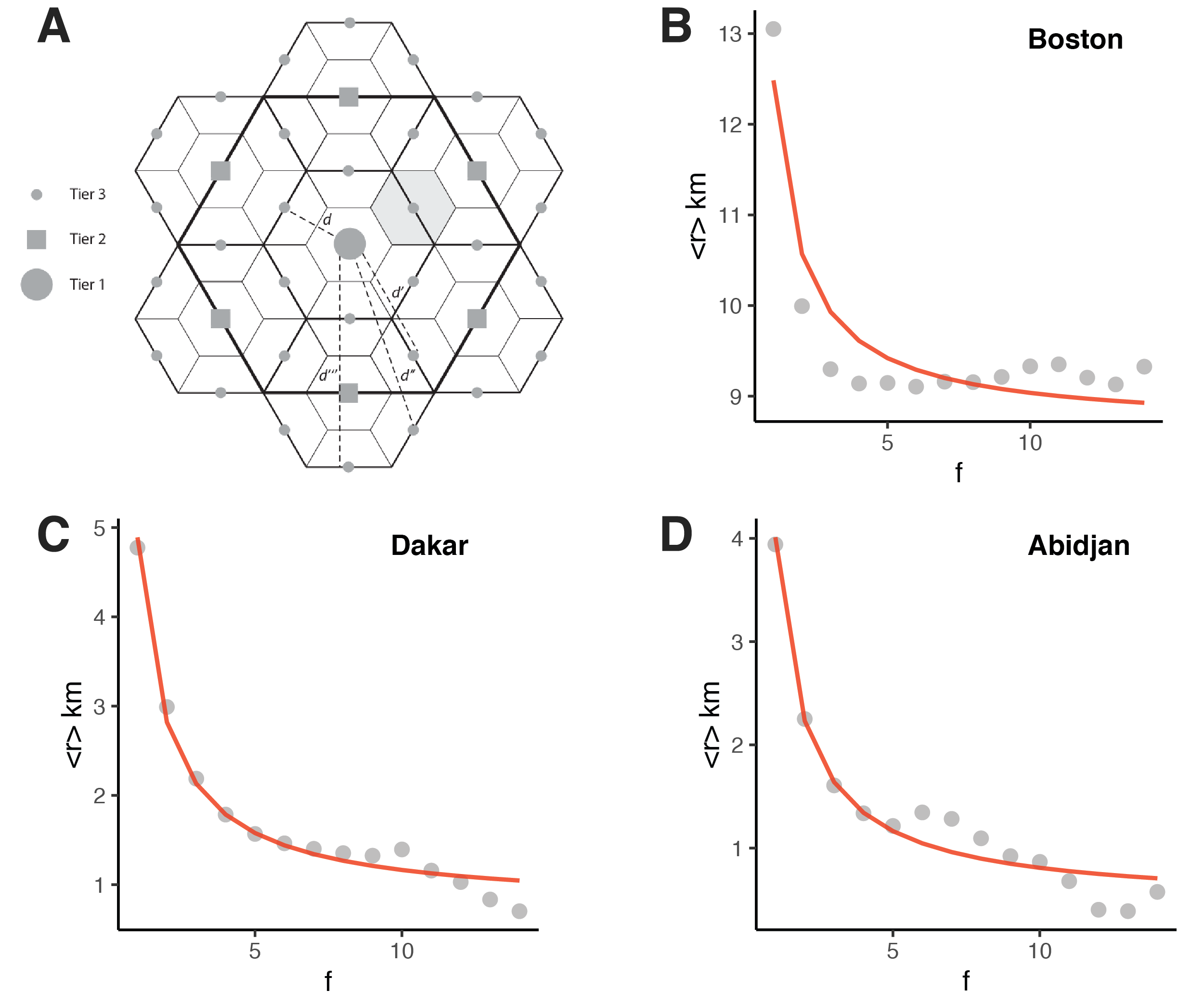}
    %\caption{}
    \begin{flushleft}
    \textbf{Fig. S5 $|$ $\langle r \rangle = K/f$ and the Central Place Theory.} \textbf{A}, Schematic figure of the Central Place Theory. The spatial arrangement of three Tiers of centers in a two dimentional space. This hierarchical arrangement of central places results in the most efficient transport network. \textbf{C-D}, The distance travelled per visit to perform activities with visiting-frequency $f$ averaged over all the individuals: Boston (B), Dakar (C), and Abidjan (D). $\langle r \rangle = K/f$ curve fits very well with the empirical observation and supports the notion of universal travel-budget.
    \end{flushleft}
    \label{fig:sfig6}
\end{figure*}

\clearpage
\begin{figure*}[!ht]
    \centering
    \includegraphics[width=.95\linewidth]{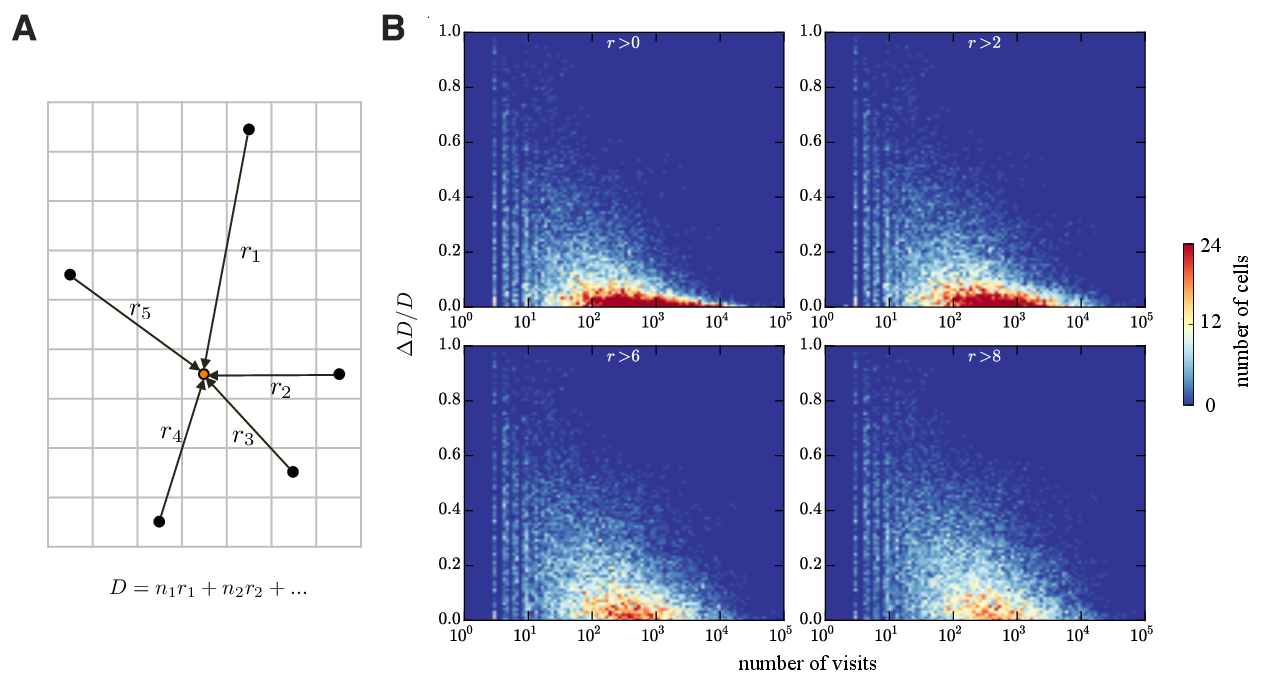}
    %\caption{}
    \begin{flushleft}
    \textbf{Fig. S6 $|$ Fermat-Torricelli-Weber (FTW) efficiency of collective human movements.}. {\bf A}) The schematic figure shows how the FTW efficiency is computed. The total distance travelled by visitors of a specific cell (red dot) can be minimized by moving the destination cell on the grid. The efficiency is $\Delta{\mathcal{D}_{total}}/{\mathcal{D}_{total}}$, which is the ratio between $\Delta{\mathcal{D}_{total}}$, i.e., the improvement gained in reducing the aggregate distance travelled by moving the cell from its actual location to the optimum FTW point, and the actual aggregate distance travelled by visitors to that cell, ${\mathcal{D}_{total}}$. {\bf B}) Each density plot represents the number of cells with a particular number of visits and FTW efficiency for Greater Boston Area based on CDR for the month of August 2009. The FTW efficiency is computed for each cell based on visits made by visitors who live at distances larger than $r_c$. These plots compare how density changes by increasing $r_c$ from 0 to $10~km$. For $r_c=0$ the density is particularly high where the FTW efficiency is very high. As the number of visits is increased, the distribution becomes narrower and the FTW efficiency increases. This pattern still survives but becomes weaker as $r_c$ is increased as described in Supplementary Material. 
    \end{flushleft}
    \label{fig:sfig7}
\end{figure*}

\begin{figure*}[!ht]
    \centering
    \includegraphics[width=.85\linewidth]{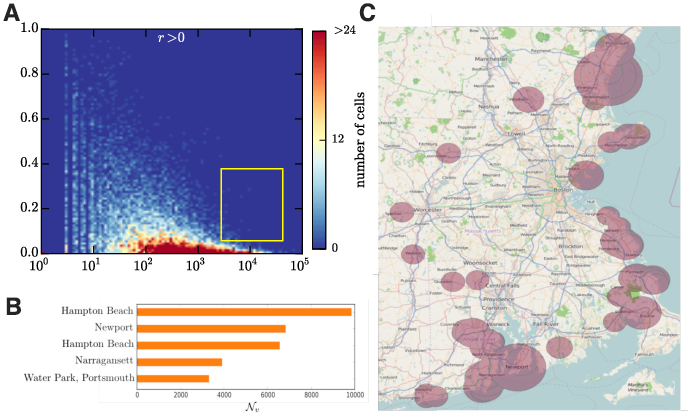}
    \begin{flushleft}
    \textbf{Fig. S7 $|$ {Transportation optimality of centers and tourism outliers for Greater Boston Area based on CDR.}} {\bf A,} The fractional distance improvement versus cell's number of visits. The higher the number of visitors, the higher is the chance that the cell is transportation efficient.  As shown in {\bf B, C}, the outliers to this pattern (the cells corresponding to points in the yellow box in (a)) are located near shores, lakes, etc., and are well-known touristic locations.
    \end{flushleft}
    \label{fig:sfig8}
\end{figure*}

\begin{figure*}[!ht]
    \centering
    \includegraphics[width=.85\linewidth]{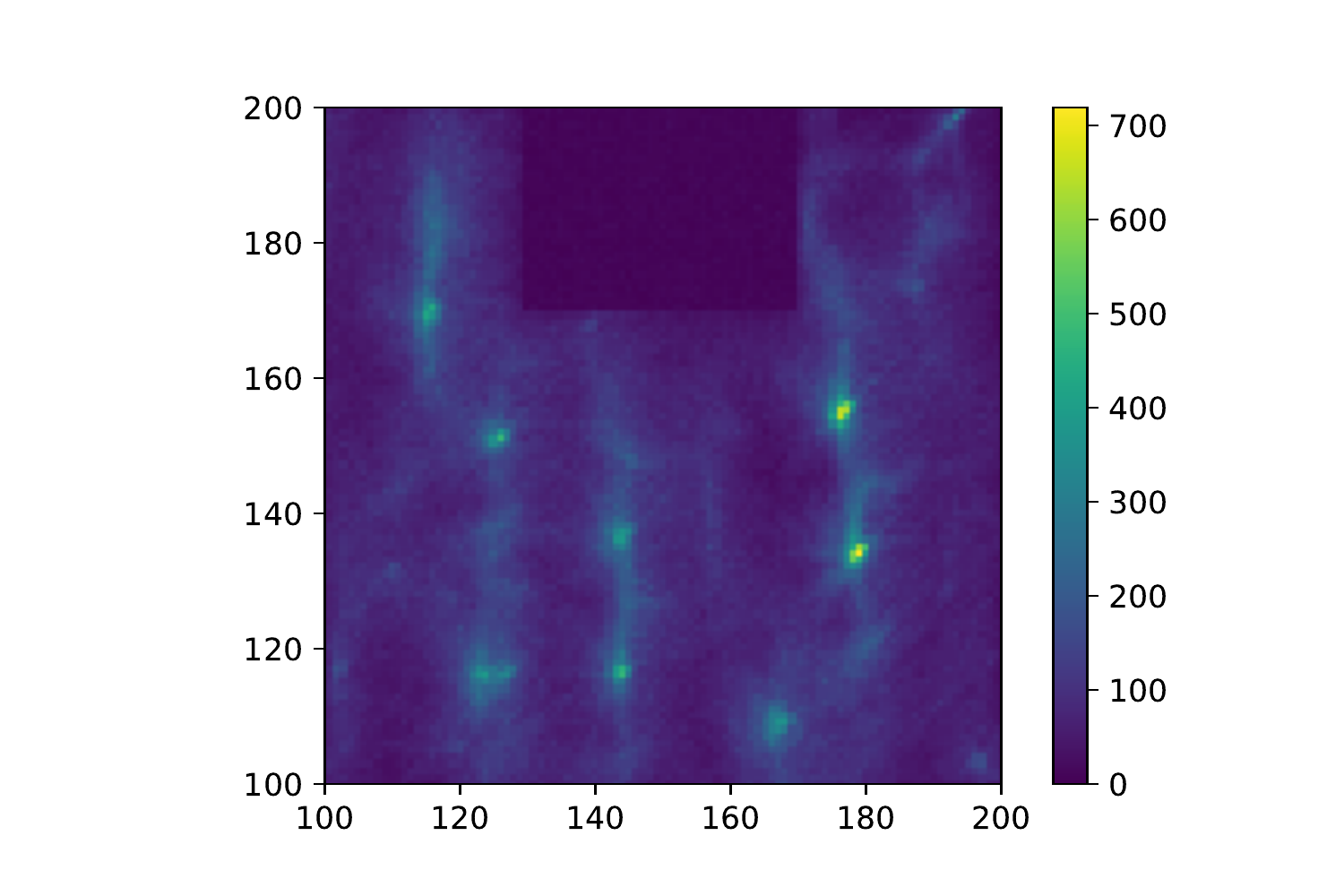}
    \begin{flushleft}
    \textbf{Fig. S8 $|$ {Visitation pattern of PEPR model on non-square lattice.}}
    \label{fig:sfig9}
    \end{flushleft}
\end{figure*}


\begin{thebibliography}{10}

\bibitem{zipf1946p}
G.~K. Zipf, {\it American Sociological Review\/} {\bf 11}, 677 (1946).

\bibitem{erlander1990gravity}
S.~Erlander, N.~F. Stewart, {\it The gravity model in transportation analysis:
  theory and extensions\/}, vol.~3 (Vsp, 1990).

\bibitem{brockmann2006scaling}
D.~Brockmann, L.~Hufnagel, T.~Geisel, {\it Nature\/} {\bf 439}, 462 (2006).

\bibitem{gonzalez2008understanding}
M.~C. Gonzalez, C.~A. Hidalgo, A.-L. Barab{\'a}si, {\it Nature\/} {\bf 453},
  779 (2008).

\bibitem{song2010modelling}
C.~Song, T.~Koren, P.~Wang, A.-L. Barab{\'a}si, {\it Nature Physics\/} {\bf 6},
  818 (2010).

\bibitem{simini2012universal}
F.~Simini, M.~C. Gonz{\'a}lez, A.~Maritan, A.-L. Barab{\'a}si, {\it Nature\/}
  {\bf 484}, 96 (2012).

\bibitem{schlapfer2020hidden}
M.~Schl{\"a}pfer, M.~Szell, Salat, C.~Ratti, G.~West, {\it
  arXiv preprint arXiv:2002.06070
\/}
  (2020).

\bibitem{mokhtarian2004ttb}
P.~L. Mokhtarian, C.~Chen, {\it Transportation Research Part A: Policy and
  Practice\/} {\bf 38}, 643 (2004).

\bibitem{christaller1933zentralen}
W.~Christaller, {\it Die zentralen Orte in S{\"u}ddeutschland\/} (Jena: Gustav
  Fischer, 1933).

\bibitem{fujita2001spatial}
M.~Fujita, P.~R. Krugman, A.~J. Venables, {\it The Spatial Economy: Cities,
  Regions, and International Trade\/} (MIT Press, 2001).

\bibitem{zipf1949human}
G.~K. Zipf, {\it Human Behavior and the Principle of Least Effort\/}
  (Addison-Wesley, 1949).

\bibitem{anas1998urban}
A.~Anas, R.~Arnott, K.~A. Small, {\it Journal of Economic Literature\/} {\bf
  36}, 1426 (1998).

\bibitem{batty2008size}
M.~Batty, {\it Science\/} {\bf 319}, 769 (2008).

\bibitem{henderson2004handbook}
V.~Henderson, J.-F. Thisse, {\it Handbook of Regional and Urban Economics:
  Cities and Geography\/}, vol.~4 (Elsevier, 2004).

\bibitem{bertaud2018order}
A.~Bertaud, {\it Order Without Design: How Markets Shape Cities\/} (MIT Press,
  2018).

\bibitem{louail2014mobile}
T.~Louail, {\it et~al.\/}, {\it Scientific Reports\/} {\bf 4}, 5276 (2014).

\bibitem{zhong2017revealing}
C.~Zhong, {\it et~al.\/}, {\it Urban Studies\/} {\bf 54}, 437 (2017).

\bibitem{rozenfeld2008laws}
H.~D. Rozenfeld, {\it et~al.\/}, {\it Proceedings of the National Academy of
  Sciences\/} {\bf 105}, 18702 (2008).

\bibitem{west2017scale}
G.~B. West, {\it Scale: the Universal Laws of Growth, Innovation,
  Sustainability, and the Pace of Life in Organisms, Cities, Economies, and
  Companies\/} (Penguin, 2017).

\bibitem{zaburdaev2015levy}
V.~Zaburdaev, S.~Denisov, J.~Klafter, {\it Reviews of Modern Physics\/} {\bf
  87}, 483 (2015).

\bibitem{strandburg2015shared}
A.~Strandburg-Peshkin, D.~R. Farine, I.~D. Couzin, M.~C. Crofoot, {\it
  Science\/} {\bf 348}, 1358 (2015).

\bibitem{batty2013new}
M.~Batty, {\it The New Science of Cities\/} (MIT Press, 2013).

\bibitem{weber1929alfred}
A.~Weber, C.~J. Friedrich, {\it et~al.\/}  (1929).

\bibitem{kreps1990game}
D.~M. Kreps, {\it Game Theory and Economic Modelling\/} (Oxford University
  Press, 1990).

\bibitem{myerson2013game}
R.~B. Myerson, {\it Game Theory\/} (Harvard University Press, 2013).

\bibitem{brubaker2013limits}
R.~Brubaker, {\it The Limits of Rationality\/} (Routledge, 2013).

\bibitem{de2014d4d}
Y.-A. de~Montjoye, Z.~Smoreda, R.~Trinquart, C.~Ziemlicki, V.~D. Blondel, {\it
  arXiv preprint arXiv:1407.4885\/}  (2014).

\bibitem{blondel2012data}
V.~D. Blondel, {\it et~al.\/}, {\it arXiv preprint arXiv:1210.0137\/}  (2012).

\bibitem{rozenfeld2011area}
H.~D. Rozenfeld, D.~Rybski, X.~Gabaix, H.~A. Makse, {\it American Economic
  Review\/} {\bf 101}, 2205 (2011).

\bibitem{cao2019quantifying}
W.~Cao, L.~Dong, L.~Wu, Y.~Liu, {\it arXiv preprint arXiv:1910.12593\/}
  (2019).

\bibitem{barbosa2018human}
H.~Barbosa, {\it et~al.\/}, {\it Physics Reports\/} {\bf 734}, 1 (2018).

\end{thebibliography}
\end{document}